\documentclass[12pt]{iopart}
\usepackage{iopams}  
\usepackage[dvips]{graphicx,color}
\usepackage{hyperref}  
\newcommand{\mylsim}{\raise0.3ex\hbox{$<$\kern-0.75em\raise-1.1ex\hbox{$\sim$}}}

\begin{document}

\title[Hadronic fluctuations at zero and non-zero chemical potential]{
Baryon number, strangeness and electric charge fluctuations
at zero and non-zero chemical potential
}

\author{C. Schmidt (for the RBC-Bielefeld Collaboration)}

\address{
Universit\"at Bielefeld, Fakult\"at f\"ur Physik\\
Postfach 100131, D-33501 Bielefeld, Germany
}
\ead{schmidt@physik.uni-bielefeld.de}
\begin{abstract}
We present results on baryon number, strangeness
and electric charge fluctuations in QCD at non-zero density and temperature
obtained from lattice calculations with almost physical quark masses. 
At vanishing chemical potential, i.e.  under conditions almost
realized at RHIC and the LHC, quartic fluctuations of net baryon number and
strangeness are large in a narrow temperature interval characterizing the
transition region from the low to the high temperature phase. Our results are
based on Taylor expansions in light and strange quark chemical potentials, i.e.
we rigorously compute corrections to bulk thermodynamic quantities at non
vanishing chemical potential, by performing a Taylor expansion in $\mu/T$. 
We find non-monotonic behavior for the radius of convergence of this series, 
which could be a hint for a critical end-point in the ($T,\mu$)-plane.
\end{abstract}



\section{Introduction}
\vspace*{-14cm}
\begin{minipage}{\textwidth}
\begin{flushright}
\texttt{\footnotesize
BI-TP 2008/09
}
\end{flushright}
\end{minipage}\\[13.5cm]
At non-zero chemical potential, lattice QCD is harmed by the ``sign-problem'', 
which makes direct lattice calculations at non-zero density practically impossible by 
all known lattice methods. Some methods, however, allow to extract 
information on the dependence of thermodynamic quantities on a small chemical 
potential parameter, based on lattice calculations performed at zero or 
imaginary chemical potential. For an overview see, e.g. \cite{overview}.

We will report here preliminary results on the Taylor expansion method with
almost realistic quark masses, i.e. a realistic strange quark mass and a light
quark mass which is about a factor of 2 heavier than the physical quark mass. The
analysis is based on data generated in the context of the recent equation of
state calculation of the RBC-Bielefeld Collaboration \cite{EoS}. The
logarithm of the grand canonical partition function or equivalently the 
pressure ($p$) can be Taylor expanded in $\mu_u/T$, $\mu_d/T$ and $\mu_s/T$ where 
$\mu_{u,d,s}$ are the up-, down- and strange-quark chemical potentials,
respectively, and $T$ is the temperature. We define
\begin{equation}
\frac{p}{T^{4}}
=\sum_{i,j,k}c^{u,d,s}_{i,j,k}(T)\left(\frac{\mu_{u}}{T}\right)^{i}
\left(\frac{\mu_{d}}{T}\right)^{j}\left(\frac{\mu_{s}}{T}\right)^{k}.
\label{eq:PTaylor}
\end{equation}
The expansion coefficients $c^{u,d,s}_{i,j,k}(T)$ 
are computed on the lattice at zero chemical potential, using  
stochastic estimators. For details see \cite{details}.

\section{Quadratic and quartic fluctuations}
The QCD partition function is naturally formulated in terms of quark-fields, and 
thus quark chemical potentials. However, alternatively to the quark chemical
potentials one can introduce chemical potentials for the conserved quantities
baryon number $B$, electric charge $Q$ and strangeness $S$ ($\mu_{B,Q,S}$), which are
related to $\mu_{u,d,s}$ via
\begin{equation}
\mu_u=\frac{1}{3}\mu_{B} +\frac{2}{3}
\mu_{Q},\qquad
\mu_{d}=\frac{1}{3}\mu_{B}-\frac{1}{3}\mu_{Q},\qquad
\mu_{s}=\frac{1}{3}\mu_{B}-\frac{1}{3}\mu_{Q}-\mu_{S}.
\label{eq:chempot}
\end{equation}
By means of these relations the coefficients $c^{B,Q,S}_{i,j,k}$ of the pressure
expansion in terms of $\mu_{B,Q,S}$ are easily
obtained, in analogy to Eq.~\ref{eq:PTaylor}
\begin{equation}
\frac{p}{T^{4}}
=\sum_{i,j,k}c^{B,Q,S}_{i,j,k}(T)\left(\frac{\mu_{B}}{T}\right)^{i}
\left(\frac{\mu_{Q}}{T}\right)^{j}\left(\frac{\mu_{S}}{T}\right)^{k}.
\label{eq:PTaylor_hadronic}
\end{equation}
The quadratic (Gaussian) fluctuations ($\chi^{B,Q,S}_2$) of $B$, $Q$ and $S$
respectively are  related by the fluctuation dissipation theorem to the second 
derivatives of the partition function with respect to the corresponding chemical 
potentials $\mu_{B,S,Q}$, whereas the quartic (non-Gaussian) fluctuations 
($\chi^{B,Q,S}_4$) are defined by the fourth derivatives. In terms of the
expansion coefficients we find at zero chemical potential
\begin{eqnarray}
\frac{\chi^{B}_2}{T^2}\equiv \frac{1}{VT^3} \left<B^2\right>=2c_{2,0,0}^{B,Q,S}; &\quad& 
{\chi^{B}_4}\equiv \frac{1}{VT} \left(\left<B^4\right>-3\left<B^2\right>^2\right)=24c_{4,0,0}^{B,Q,S}; 
\nonumber \\
\frac{\chi^{Q}_2}{T^2}\equiv \frac{1}{VT^3} \left<Q^2\right>=2c_{0,2,0}^{B,Q,S}; &\quad&
{\chi^{Q}_4}\equiv \frac{1}{VT} \left(\left<Q^4\right>-3\left<Q^2\right>^2\right)=24c_{0,4,0}^{B,Q,S}; 
\nonumber \\
\frac{\chi^{S}_2}{T^2}\equiv \frac{1}{VT^3} \left<S^2\right>=2c_{0,0,2}^{B,Q,S}; &\quad&
{\chi^{S}_4}\equiv \frac{1}{VT} \left(\left<S^4\right>-3\left<S^2\right>^2\right)=24c_{0,0,4}^{B,Q,S}. 
\nonumber 
\end{eqnarray}

\begin{figure}
\includegraphics[width=0.49\textwidth]{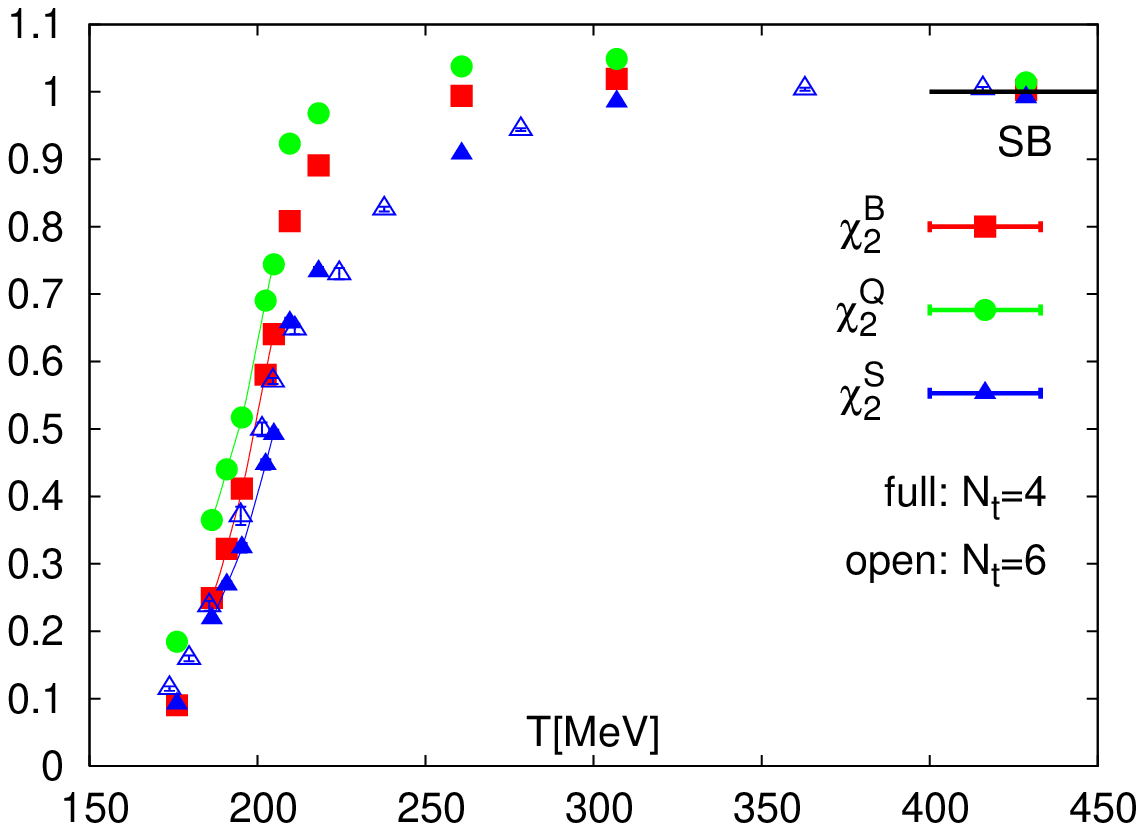}
\includegraphics[width=0.49\textwidth]{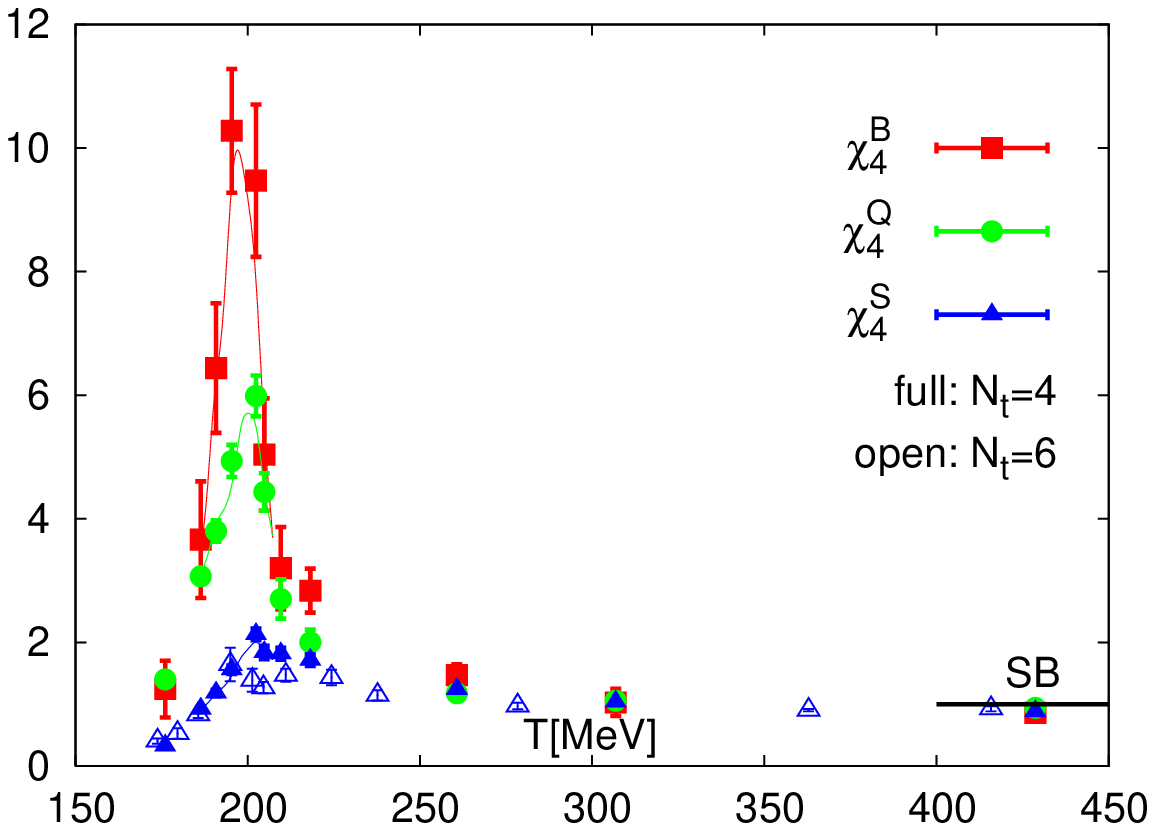}
\caption{Quadratic (left) and quartic fluctuation (right) of baryon number $(B)$, electric
charge $(Q)$ and strangeness $(S)$, normalized by their corresponding Stefan-Boltzmann value.
The results are from $N_t=4$ lattices, and in the case of strangeness fluctuations, open 
symbols show results from $N_t=6$ lattices. Thin solid lines are from Ferrenberg-Swendsen analyses.}
\label{fig:c2c4}
\end{figure}

In Fig.~\ref{fig:c2c4} we show these fluctuations as a function of the temperature. 
The quadratic fluctuations rapidly increase at a transiton temperature $T=T_c$ and quickly 
approach 1 for $T>T_c$.
Note, that all fluctuations have been normalized by their corresponding Stefan-Boltzmann 
value. The quartic fluctuations show a peak at $T=T_c$, which is most pronounced for the 
baryon number fluctuations and least pronounced for the strangeness fluctuations. In fact,
by using an appropriate scaling Ansatz for the free energy, one can show that $\chi_4$
will develop a cusp in the chiral limit, where the transition becomes 2nd order.
For the strangeness fluctuations we show results from two different lattice spacings ($N_t=4,6$)
and find a relatively small cut-off dependence, which is of similar magnitude as we have found for
the pressure itself \cite{EoS}. 

\begin{figure}
\includegraphics[width=0.49\textwidth]{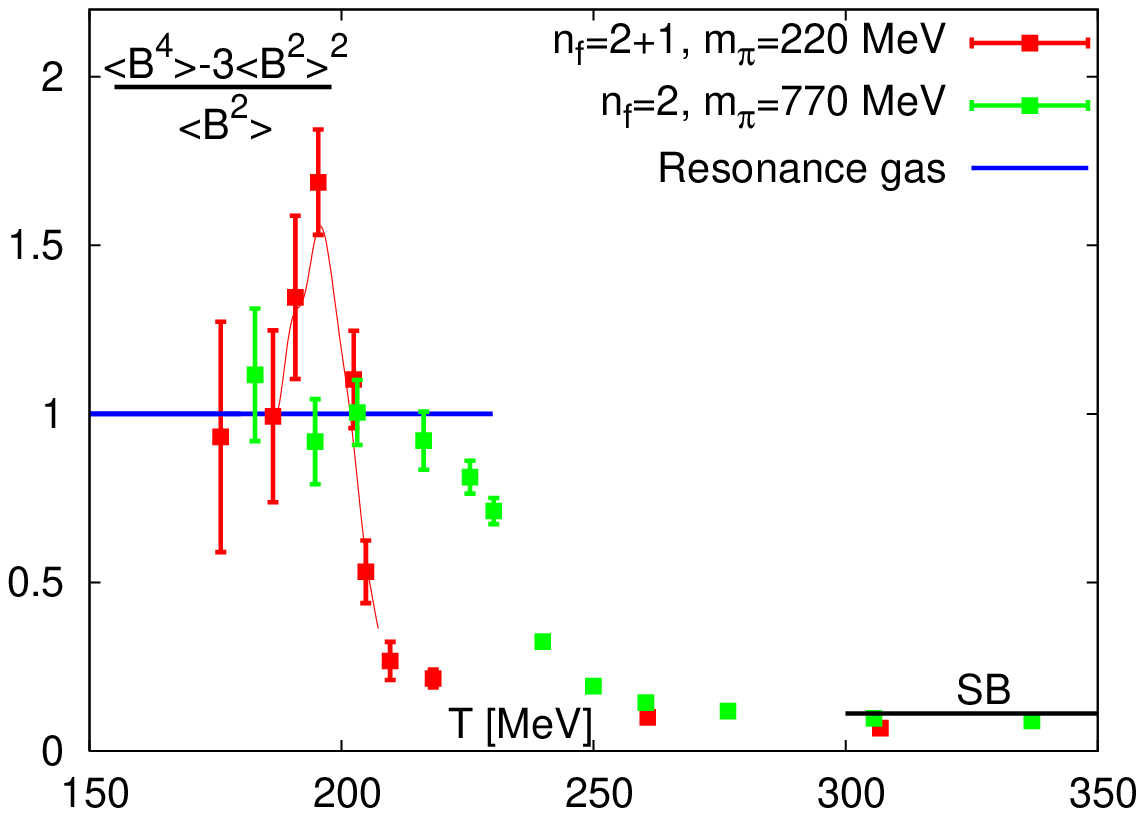}
\includegraphics[width=0.49\textwidth]{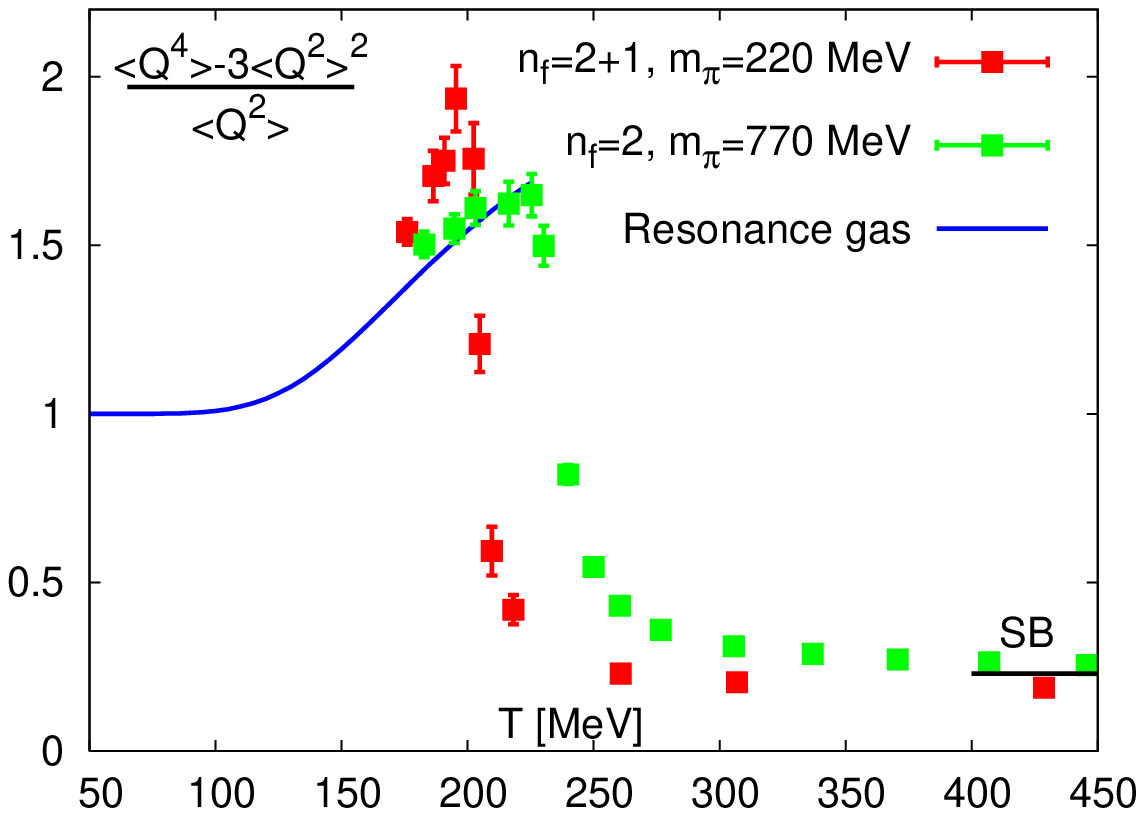}
\caption{Ratio of quartic and quadratic fluctuations for baryon number $B$ (left) and electric
charge $Q$ (right). Shown are the preliminary results for 2+1-flavor QCD with $m_{\pi}\approx 220$~MeV and 
the previous results for 2-flavor with $m_{\pi}\approx 770$~MeV. Thin solid line (left) is a 
Ferrenberg-Swendsen analysis.}
\label{fig:ratio}
\end{figure}

In Fig.~\ref{fig:ratio} we show the ratios of $\chi_2$ and $\chi_4$, for fluctuations of $B$ and $Q$.
Ratios are well suited quantities to compare with experiment, since many systematic
errors as well as the volume dependence are eliminated by taking the ratio.
In Fig.~\ref{fig:ratio}, we also compare our preliminary (2+1)-flavor data with a pion mass of 
$m_\pi\approx220$~MeV to the previous 2-flavor lattice results with $m_\pi\approx770$~MeV \cite{pre},
as well as the resonance gas predictions (for $T<T_c$). Besides a smaller $T_c$, we find that
the fluctuations now significantly rise above the resonance gas level, which was previously not
observed. Note also, that this quantity directly gives access to the relevant degrees of freedom, 
thus already above $1.5 T_c$ the quantum numbers of the medium are those of a quasi free gas of 
quarks and gluons \cite{pre}.   

\section{The radius of convergence and the phase diagram}
\begin{figure}
\begin{center}
\includegraphics[width=0.49\textwidth]{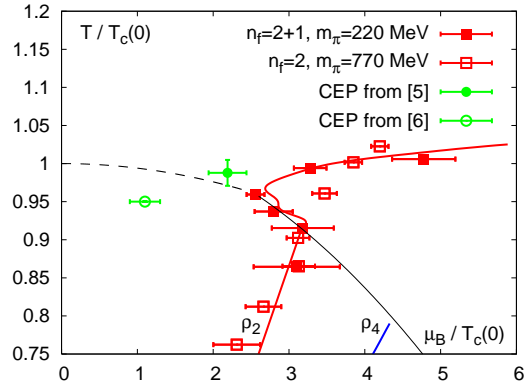}
\end{center}
\caption{The approximation $\rho_2$ of the radius of convergence of the Taylor series of the pressure with 
respect to the baryon chemical potential, as explained in the text. Shown are the preliminary results for 
2+1-flavor with $m_{\pi}\approx 220$~MeV and the previous results for 2-flavor with $m_{\pi}\approx 770$~MeV,
as well as the resonance gas values for $\rho_2, \rho_4$ and earlier calculations of the CEP [5, 6]. 
Thin dashed and solid lines indicate a suggested phase diagram and are only meant to guide the eye.}
\label{fig:radius}
\end{figure}

Estimating the radius of convergence of the series (Eq.~\ref{eq:PTaylor_hadronic}) is a method 
to locate the critical end-point (CEP) in the $(T,\mu_B)$-phase diagram. 
The radius of convergence ($\rho$) is defined as
\begin{equation}
\rho = \lim_{n\to\infty}\rho_n\qquad\mbox{with}\qquad\rho_n=\sqrt{{c^{B,Q,S}_{n,0,0}}/{c^{B,Q,S}_{n+2,0,0}}}
\end{equation}  
For $T>T_{CEP}$, the estimators 
for the radius of convergence should be large and eventually will diverge.
For $T<T_{CEP}$ the radius of convergence will be limited by the phase transition line. 
In Fig.~\ref{fig:radius} we plot the first approximation for the radius of convergence $\rho_2$ in the 
$(T,\mu_B)$-plane, together with previous results for $\rho_2$ \cite{pre}, as well as previous 
estimates for the CEP from \cite{FK,GG}.  
Also shown are the resonance gas values of $\rho_2$ and $\rho_4$, which seems to be aproached by the lattice
data for $T\mylsim 0.85T_c$.
For an undoubtful determination of the critical point higher approximations for
$\rho$ will be required, however the non monotonic behaviour which is now seen for $m_\pi\approx 220$ MeV
and has not been seen for $m_\pi\approx 770$ MeV might be a first sign of the critical region of the CEP.
In addition, thin dashed and solid lines indicate a suggested phase diagram and are only meant to guide the eye.

\ack
We would like to thank all members of the RBC-Bielefeld Collaboration for
helpful discussions and comments. The work has been supported in parts by the
U.S. Department of Energy under Contract No. DE-AC02-98CH10886. 
Numerical simulations have
been performed on the QCDOC computer of the RIKEN-BNL research center, the DOE
funded QCDOC at BNL and the apeNEXT at Bielefeld University.

\end{document}